\documentclass[preprint,aps]{revtex4}
\usepackage{graphicx,eucal,bm}
\begin{document}
\title{Conformal invariance in two-dimensional turbulence}
\author{D. Bernard$^{\dagger}$, G. Boffetta$^{\star}$,  A. Celani$^{\bullet}$
and G. Falkovich$^{\&}$}
\affiliation{$^{\dagger}$ Service de Physique Th\'eorique de Saclay,
CEA/CNRS, Orme des Merisiers, 91191 Gif-sur-Yvette Cedex, France
\\$^{\star}$
Dipartimento di Fisica Generale and INFN, Universit\`a di Torino,
via Pietro Giuria 1, 10125 Torino, Italy
\\$^{\bullet}$ CNRS, INLN, 1361 Route des Lucioles,
06560 Valbonne Sophia Antipolis, France\\$^{\&}$ Physics of
Complex Systems, Weizmann Institute of Science,\\ Rehovot 76100,
Israel }
%\date{\today}
\maketitle

{\bf Simplicity of fundamental physical laws manifests itself in
fundamental symmetries. While systems with an infinity of strongly
interacting degrees of freedom (in particle physics and critical
phenomena) are hard to describe, they often demonstrate
symmetries, in particular scale invariance. In two dimensions (2d)
locality often promotes scale invariance to a wider class of
conformal transformations which allow for nonuniform re-scaling.
Conformal invariance allows a thorough classification of
universality classes of critical phenomena in 2d. Is there
conformal invariance in 2d turbulence, a paradigmatic example of
strongly-interacting non-equilibrium system? Here, using numerical
experiment, we show that some features of 2d inverse turbulent
cascade display conformal invariance. We observe that the
statistics of vorticity clusters is remarkably close to that of
critical percolation, one of the simplest universality classes of
critical phenomena. These results represent a new step in the
unification of 2d physics within the framework of conformal
symmetry.}

%Even more remarkably, often (but not always \cite{RC}) scale
%invariance is local so that the statistical properties are actually
%invariant with respect to a much wider class of transformations: this
%is the conformal invariance \cite{Polyakov}.

We consider here 2d incompressible turbulent motion of a fluid,
which represents an appropriate description of large-scale motions
of the atmosphere and can be realized in different laboratory
settings as well$^{1-5}$. As predicted by Kraichnan$^1$, stirring
at some scale $L_f$ results in two turbulence cascades, with the
formation of fine-scale vortical structures and large-scale
velocity structures. In two dimensions, squared vorticity
$\omega^2=({\bm \nabla} \times {\bm v})^2$ performs a direct
cascade to small scales while kinetic energy $\frac{1}{2}{\bm
v}^2$ flows from the injection length $L_f$ to large scales,
opposite to the three-dimensional case.
%Note that the kinetic energy and all powers of vorticity are inviscid
%invariants of incompressible hydrodynamics.
We focus here on the inverse cascade of energy for which, not
surprisingly in view of the presence of a strong interaction,
there is no exact analytic theory.  Phenomenological dimensional
arguments give consistent predictions, though in two seemingly
unrelated ways.  Consider the velocity difference $v_r$ at the
distance $r$. On the one hand, one may require that the kinetic
energy $v_r^2$ divided by the typical time $r/v_r$ must be
constant and equal to the energy flux, $\epsilon$: $v_r^3\sim
\epsilon r$. On the other hand, it can be argued that vorticity,
which cascades to small scales, must be in equipartition in the
inverse cascade range$^6$. If this is the case, the enstrophy
$r^d\omega_r^2$ accumulated in a volume of size $r$ is
proportional to the typical time $r/v_r$ at such scale, i.e.
$r^d\omega_r^2\sim r/v_r$.
%If this is the case, the pair correlation function at
%scale $r$ is proportional to the typical time $r/v_r$ spent
%at such scale, i.e.$\omega_r^2\sim r^{1-d}/v_r$ \cite{FGV}.
Using $\omega_r\sim v_r/r$ we derive $v_r^3\sim r^{3-d}$ which for
$d=2$ is exactly the requirement of constant energy flux.
Amazingly, the requirements of vorticity equipartition (i.e.
equilibrium) and energy flux (i.e. turbulence) give the same
Kolmogorov-Kraichnan scaling in 2d. Experiments$^{4,5,7}$ and
numerical simulations$^{8}$ indeed demonstrate scale-invariant
statistics with the vorticity having scaling dimension $2/3$:
$\omega_r\propto r^{-2/3}$.

Our goal here is to find out whether scale invariance can be
extended to conformal invariance at least for some properties of
2d turbulence. Under conformal transformations the
  lengths are re-scaled non-uniformly yet the angles between vectors
  are left unchanged (a useful property in navigation cartography
  where it is often more important to aim in the right direction than
  to know the distance)$^{9,10}$. The novelty of our approach is that we analyze the
inverse cascade by describing the large-scale statistics of the
boundaries of vorticity clusters, i.e. large-scale zero-vorticity
lines. In equilibrium critical phenomena, cluster boundaries in
the continuous limit of vanishingly small
lattice size %and at criticality
were recently found  to belong to a remarkable class of curves
that can be mapped into Brownian walk (called Stochastic Loewner
Evolution or SLE curves)$^{11-19}$. Namely, consider a curve
$\gamma(t)$ that starts at a point on the boundary of the
half-plane $H$ (by conformal invariance any planar domain is
equivalent to the upper half plane). One can map the half-plane
$H$ minus the curve $\gamma(t)$ back onto $H$ by an analytic
function $g_t(z)$ which is unique upon imposing the condition
$g_t(z)\sim z+2t/z+{\rm O}(1/z^2)$ at infinity. The growing tip of
the curve is mapped into a real point $\xi(t)$.
Loewner$^{20}$ found in 1923 %\cite{Loew}
that %this family of conformal maps satisfies a simple evolution equation so that
the conformal map $g_t(z)$ and the curve $\gamma(t)$ are fully
parametrized by the driving function $\xi(t)$.
%One defines the hull $K(t)$ which is the union
%of the curve with the set of points that cannot be reached from
%infinity without crossing it. Since the complement of $K(t)$ in
%$H$ is simply connected it can be mapped
%Loewner found that the evolution of the curve $\gamma(t)$, also called
%the ``trace'' of the process, can be univocally linked to the
%evolution of the mapping upon imposing the condition $g_t(z)\sim
%z+2t/z+{\rm o}(1/z)$ at $z\to\infty$. The family of conformal maps
%$g_t(z)$ satisfies the Loewner equation
%$dg_t(z)/dt=2[g_t(z)-\xi(t)]^{-1}$ \cite{Loew}.
Almost eighty
years later, Schramm$^{11}$ considered random curves in planar
domains and showed that their statistics is conformal invariant if
$\xi(t)$ is a Brownian walk, i.e. its increments are identically
and independently distributed and
$\langle(\xi(t)-\xi(0))^2\rangle=\kappa t$.
%then the statistics of $\gamma(t)$ is conformally invariant \cite{Schramm}.
In simple words, the locality in time of the Brownian walk
translates into the local scale-invariance of SLE curves, i.e.
conformal invariance.  SLE$_\kappa$ provide a natural
classification (by the value of the diffusivity $\kappa$) of
boundaries of clusters of 2d critical phenomena$^{16}$ described
by conformal field theories (CFT)$^{10}$ and allow to establish
many new results (see \cite{Lbook,GK,Cardy,BB1} for a review).

\begin{figure}[!h]
\centerline{\includegraphics[width=16cm]{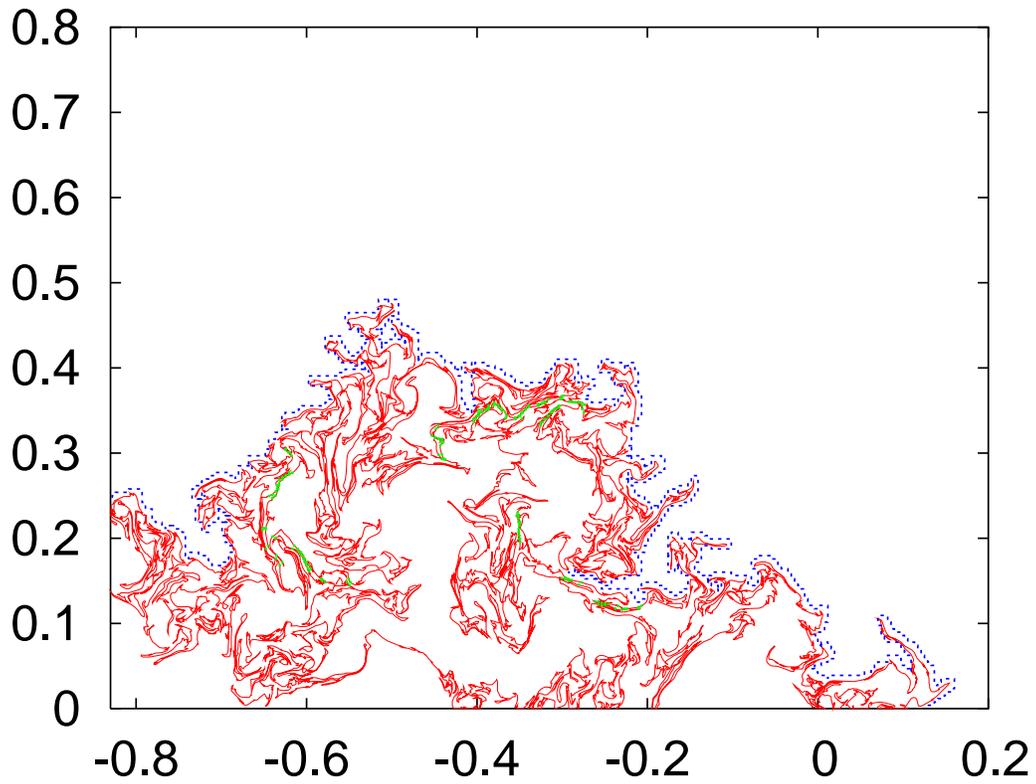}} \caption{A
portion of a candidate SLE trace obtained from the vorticity
field. The red curve is a zero-vorticity line in the upper
half-plane. The dashed blue line is the "outer boundary" of the
red curve, i.e. the boundary of the region that can be reached
from infinity without getting closer than  $L_f$ to the red curve.
The green dots mark the necks of large fjords and peninsulae.}
\label{fig:1}
\end{figure}

The fractal dimension of SLE$_\kappa$ curves is known to
be$^{22,23}$ $D_\kappa=1+\kappa/8$ for $\kappa<8$. To establish
possible link, let us try to relate the Kolmogorov-Kraichnan
phenomenology to the fractal dimension of the boundaries of
vorticity clusters. Note that one ought to distinguish between the
dimensionality $2$ of the full vorticity level set (which is
space-filling) and a single zero-vorticity line that encloses a
large-scale cluster$^{24}$. Consider the vorticity cluster of
gyration radius $L$ which has the ``outer boundary''  of perimeter
$P$ (that boundary is the part of the zero-vorticity line
accessible from outside, see Fig.~\ref{fig:1} for an
illustration). The vorticity flux through the cluster, $\int
\omega dS \sim \omega_L L^2$, must be equal to the velocity
circulation along the boundary, $\Gamma=\oint {\bm v}\cdot d{\bm
\ell}$.  The Kolmogorov-Kraichnan scaling is
$\omega_L\sim\epsilon^{1/3}L^{-2/3}$ (coarse-grained vorticity
decreases with scale because contributions with opposite signs
partially cancel) so that the flux is $\propto L^{4/3}$. As for
circulation, since the boundary turns every time it meets a
vortex, such a contour is irregular  on scales larger than the
pumping scale. Therefore, only the velocity at the pumping scale
$L_f$ is expected to contribute to the circulation, such velocity
can be estimated as $(\epsilon L_f)^{1/3} $ and it is independent
of $L$. Hence, circulation should be proportional to the
perimeter, $\Gamma\propto P$, which gives $P\propto L^{4/3}$, i.e.
the fractal dimension of the exterior of the vorticity cluster is
expected to be $4/3$.

\begin{figure}[!h]
\centerline{\includegraphics[width=14cm]{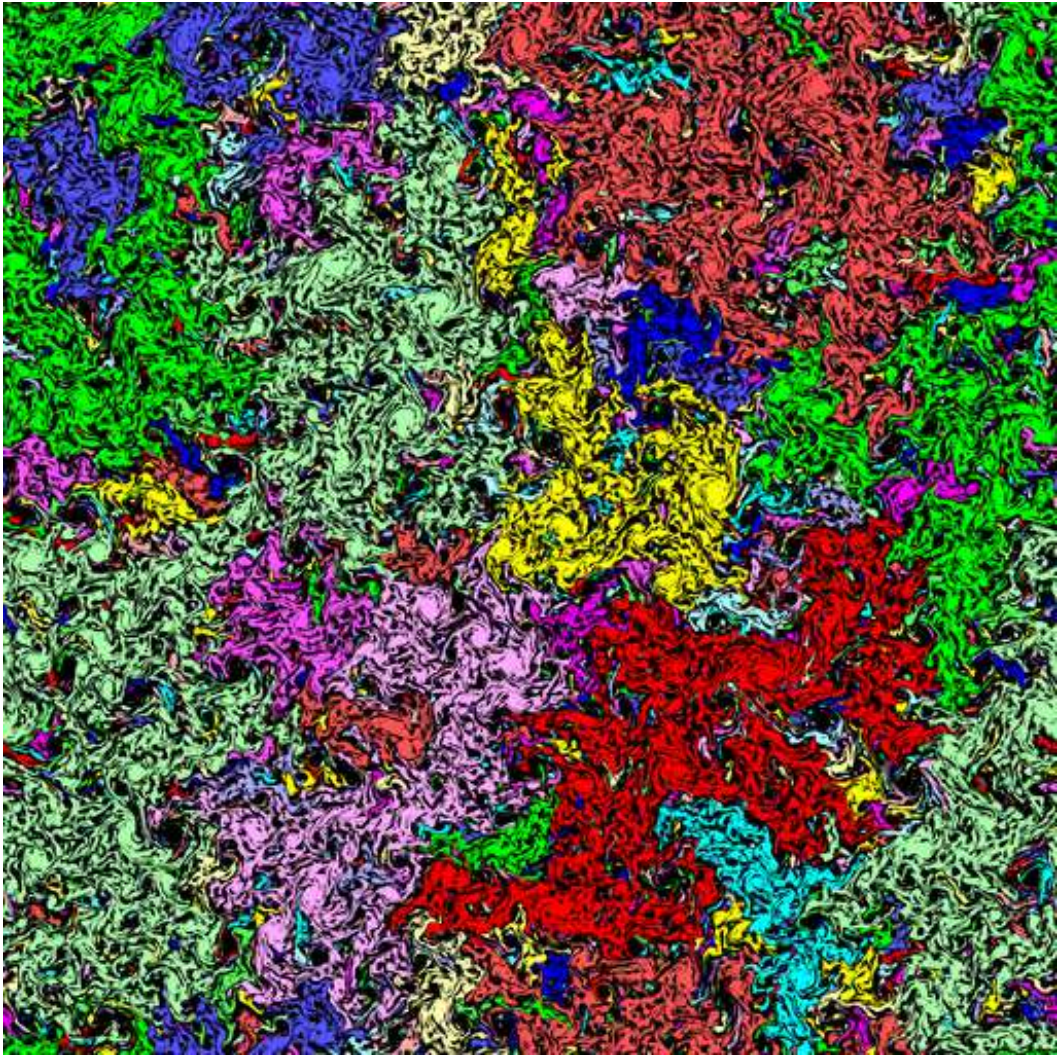}}
\caption{Vorticity clusters. These are defined as connected
regions with the same sign of vorticity (here positive). Colours
are arbitrarily attributed to different clusters. Regions of
negative vorticity are black. The forcing lengthscale $L_f$ is one
hundredth of the box side.} \label{fig:0}
\end{figure}

\begin{figure}[!h]
\centerline{\includegraphics[width=16cm]{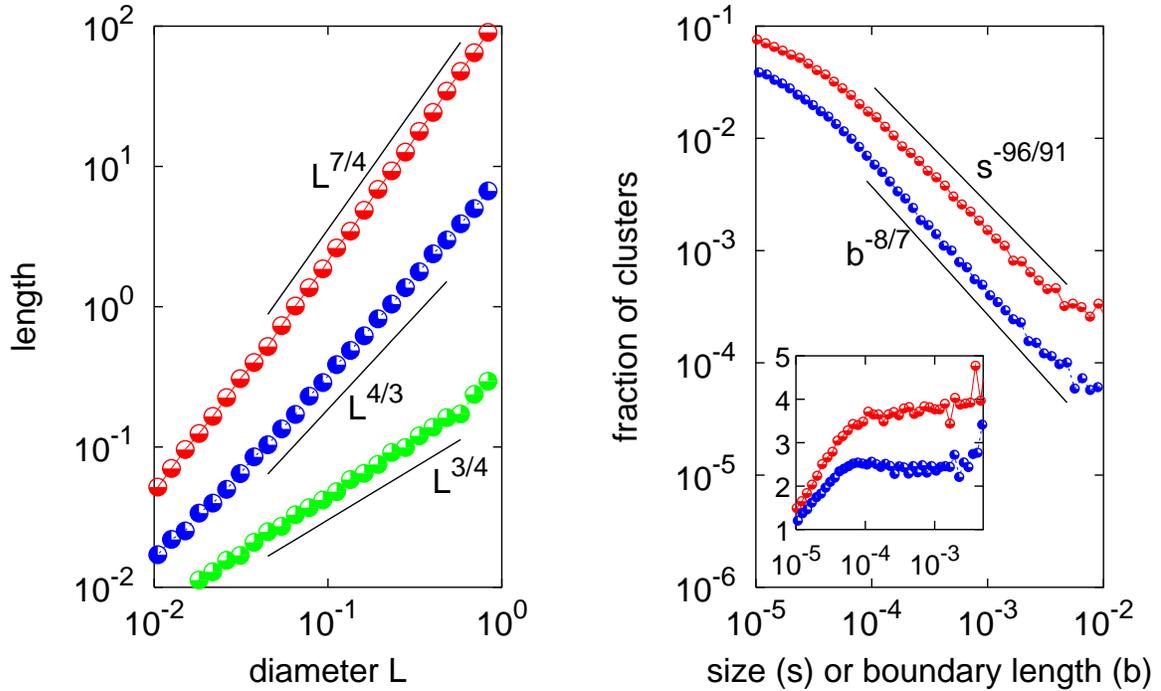}}
\caption{Fractal dimensions and probabilities of size and boundary
length for vorticity clusters. {\em Left panel}: The fractal
dimensions are  the slopes of the length-diameter dependencies in
log-log coordinates for the boundary of filled clusters (red), for
the outer boundary (blue) and for the necks of large
fjords/peninsulae (green). The solid lines have slopes with the
exact values for SLE$_6$ curves, $7/4$, $4/3$, $3/4$,
respectively. Fractal dimensions are obtained by computing the
average length for a given diameter of the cluster. {\em Right
panel:\/} The fraction of clusters with sizes between $s$ and
$1.25 s$ (red symbols) and with boundary lengthes between $b$ and
$1.25 b$. The solid lines are the predictions from the percolation
theory. Inset shows the same data multiplied by $s^{96/91}$ and
$b^{8/7}$, respectively. The vertical scale is linear to
appreciate the plateau in the compensated plot.} \label{fig:4}
\end{figure}

Let us check this hypothesis by data analysis. A powerful tool for
the study of 2d turbulence is the numerical integration of the
incompressible Navier-Stokes equations in a planar domain. By this
method it is possible to achieve a range of dynamical lengthscales
of about four decades, whereas current laboratory experiments are
limited to a scale separation of about a hundred. We present here
the analysis of very high resolution numerical simulations (with
up to 16384$^2$ grid points) of two-dimensional
inverse cascade.  (See Table~\ref{table} %and Ref.~\cite{BC}
for the details). Vorticity clusters are shown on
Figure~\ref{fig:0}. The fractal dimension of their exterior
boundary (without self-intersections) shown in the left panel of
Figure~\ref{fig:4} as indeed close to $D_*=4/3$. Moreover, the
fractal dimension of the boundary itself is close to $D=7/4$. Of
course, having some particular dimension does not by itself imply
that the curve belong
to SLE. Note, however, that the exterior perimeter of SLE$_\kappa$ %hulls
with $\kappa>4$ is conjectured$^{25}$ to look locally as
SLE$_{\kappa_*}$ curve with $\kappa_*=16/\kappa < 4$ resulting in
the duality relation, $(D-1)(D_*-1)=1/4$, as observed in our
turbulence data. Moreover, $D_*=4/3$ corresponds to a SLE$_{8/3}$
curve which represents the continuum limit of a self-avoiding
random walk while the dual SLE$_6$ curve corresponds to a cluster
boundary in critical percolation. That prompts us to compare the
probability distributions of sizes and boundary lengths between
vorticity and percolation clusters. The size $s$ of a cluster is
the number of connected sites with the same sign of vorticity, the
boundary length $b$ is the number of sites that belong to the
cluster but are adjacent to sites of different sign, and the
diameter $L$ is the side of the smallest square that covers the
cluster.  The results shown in the right panel of Fig.~\ref{fig:4}
are in a good agreement with the exact results from percolation
theory.

The two SLEs with $\kappa=6$ and $\kappa_*=8/3$ correspond to CFT
with zero central charge which means that the scale invariance
remains unbroken even when the system is on a manifold with
corners or with a nonzero Euler number (a topological invariant
determined by the number of handles and boundaries). Also, SLE$_6$
curves are singled out by a ``locality property'' (the curve does
not feel the boundary until it touches it), while their dual
SLE$_{8/3}$ has the ``restriction property'' (the statistics of
the curves conditioned not to visit some region is the same as in
the domain without this region)$^{12-16}$. How does all this
relate to two-dimensional turbulence?

\begin{figure}[!h]
\centerline{\includegraphics[width=16cm]{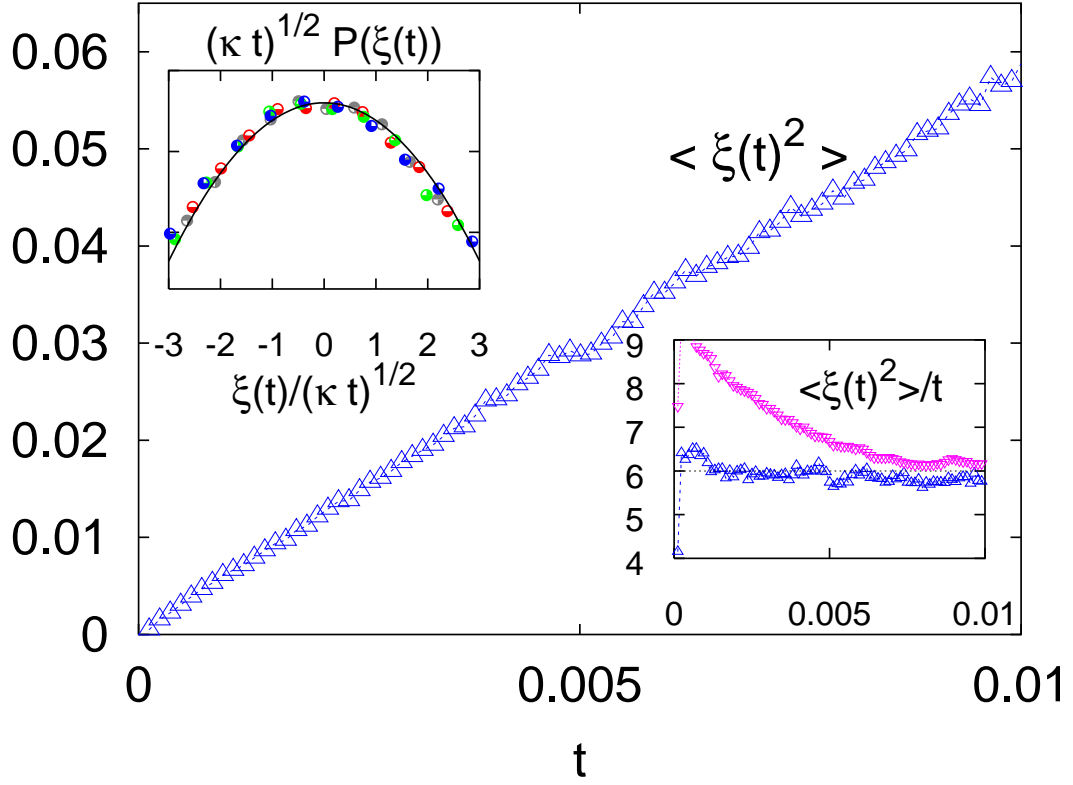}}
\caption{The driving function is an effective diffusion process
with diffusion coefficent $\kappa = 6 \pm 0.3$. The inverse
cascade range corresponds to $5\cdot10^{-5} < t < 10^{-2}$. {\em
Main frame\/}: the linear behaviour of $\langle \xi(t)^2 \rangle$.
{\em Lower-right inset \/}: Diffusivity: blue for vorticity
isolines, pink for the field with randomized phases. {\em
Upper-left inset\/}: the probability density function of the
rescaled driving function $\xi(t)/\sqrt{\kappa t}$ at four
different times $t = 0.0012, 0.003, 0.006, 0.009$; the solid line
is the Gaussian distribution $g(x)=(2 \pi)^{-1/2} \exp(-x^2/2)$.}
\label{fig:2}
\end{figure}

Now we show by a straightforward check that, within statistical
accuracy, large-scale zero-vorticity lines are indeed SLE curves,
that is they are conformal invariant and possess remarkable
properties of a kind that has never been studied in turbulence.
Zero-vorticity isolines that are candidate SLE traces are
identified as follows. First, a horizontal line representing the
real axis in the complex plane is drawn across the vorticity
field. Second, an explorer starting from the origin at the real
axis walks on the zero-vorticity isoline keeping the positive
vorticity sites always on the right. Third, when the explorer hits
the real axis it treads on it always leaving the positive region
on its right side until it can re-enter the upper half-plane. This
eventually leads the explorer to infinity in an unbounded domain.
An example of the outcome of this search is shown in
Figure~\ref{fig:1}.  Strictly speaking, this procedure faithfully
reproduces all the details of the statistics only if there is a
locality property (meaning that the exploration process does not
feel the boundary before it hits it), which holds for SLE$_6$.
Since we obtain as a result SLE with $\kappa\approx6$, our
procedure is self-consistent; we also checked that shifting and
turning the line does not modify the results presented below. To
determine which driving function $\xi(t)$ can generate such a
curve, one needs to find the sequence of conformal maps $g_t(z)$
that map the half-plane $H$ minus the curve into $H$ itself. We
approximate $g_t(z)$ by a composition of discrete, conformal slit
maps that swallow one segment of the curve at a time (a slight
variation of the techniques presented in \cite{MR}). This results
in a sequence of ``times'' $t_i$
%(equal half the capacity of the hulls)
and driving values $\xi_i$
that approximate the true driving functions.
%A small subset of driving functions is shown in Figure~\ref{fig:2}.
If the zero-vorticity isolines in the half-plane are actually SLE
traces, then the driving function should behave as an effective
diffusion process at sufficiently large times. We have collected
1,607 putative traces. The data presented by blue triangles in
Figure~\ref{fig:2} show that the ensemble average $\langle
\xi(t)^2 \rangle$ indeed grows linearly in time: the diffusion
coefficient $\kappa$ is very close to the value $6$, with an
accuracy of 5\% (see inset).  The average $\langle\xi(t)\rangle$
vanishes by the reflection symmetry of the Navier-Stokes
equations.  Additionally, the probability distribution functions
of $\xi(t)/\sqrt{\kappa t}$ collapse onto a standard Gaussian
distribution at all times $t$. Therefore, we expect the driving
$\xi(t)$ tend to a true Brownian motion and zero-vorticity lines
to become SLE$_\kappa$ traces with $\kappa$ very close to $6$ in
the limit of vanishingly small $L_f$. To appreciate how remarkable
this property is, pink symbols in the lower insert in
Fig~\ref{fig:2}  show for comparison the results of the same
procedure for the isolines of a Gaussian field having the same
Fourier spectrum as vorticity but randomized phases. The slow
incomplete recovery to $\kappa=6$ for the random-phase field
occurs at the scales where the power-law correlation is already
cut-off by friction and the field becomes truly uncorrelated.

\begin{figure}[!h]
\centerline{\includegraphics[width=16cm]{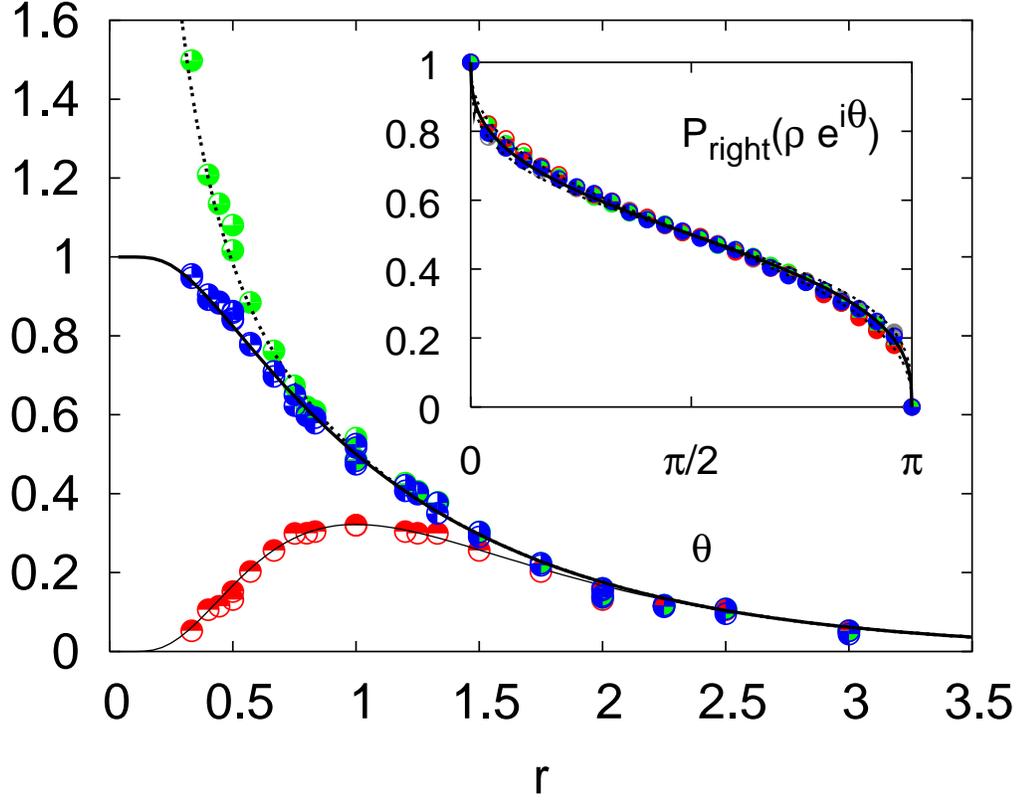}}
\caption{Crossing and surrounding probability for vorticity
clusters. {\em Main frame\/}: the probability $\pi_v$ that a
cluster crosses from top to bottom a rectangle of aspect ratio $r$
(blue), the average number $N_v$ of vertically crossing clusters
(green), and the probability $\pi_{hv}$ of a "four-legged" cluster
joining all sides of the rectangle (red). The lines are the exact
results for $\kappa=6$:
$\pi_v=\frac{3\Gamma(\frac{2}{3})}{\Gamma(\frac{1}{3})^2}
\eta^{1/3} \;{ }_2 F_1 \left(\frac{1}{3},\frac{2}{3};
\frac{4}{3};\eta \right) $ with $\eta=[(1-k)/(1+k)]^2$ and
$r=K(1-k^2)/[2K(k^2)]$ (Cardy-Smirnov, thick solid line);
$N_v=\frac{1}{2}[\pi_v+\pi_{hv}-\frac{\sqrt{3}}{2\pi}\log\eta]$
(Cardy, thick dashed line); $\pi_{hv}=\pi_v-
\frac{\eta}{\Gamma(\frac{2}{3})\Gamma(\frac{1}{3})} \;{ }_3 F_2
\left(1,1,\frac{4}{3};2, \frac{5}{3};\eta \right)$
(Watts-Dub\'edat, thin dotted line). {\em Inset\/}: the
probability that a zero-vorticity line in the upper half-plane
leaves the point $\rho e^{i \theta}$ to its right, for
$\rho=0.048,0.064,0.080,0.096$. The prediction for SLE$_\kappa$
traces is $P=\frac{1}{2}+
\frac{\Gamma(\frac{4}{\kappa})}{\sqrt{\pi}\Gamma(\frac{8-\kappa}{2\kappa})}
\;{ }_2 F_1 \left(\frac{1}{2},\frac{4}{\kappa}; \frac{3}{2};
  -\cot^2\theta\right) \cot \theta$, shown  as a thick solid line for
$\kappa=5.9$ (the best fit). The dashed lines are the
probabilities for $\kappa=5.7$ and $\kappa=6.1$.} \label{fig:3}
\end{figure}

The identification of isovorticity lines as SLE$_\kappa$ curves
allows to apply powerful techniques borrowed from the theory of
stochastic differential equations and conformal mapping theory and
to obtain analytic predictions for some nontrivial statistical
properties of vorticity clusters. The first example is the
probability that a point $z=\rho e^{i\theta}$ inside the upper
half plane is surrounded by a positive vorticity cluster connected
to the positive real axis. In this event, it is not possible to
reach infinity with a continuous path starting at $z$ without
treading on positive vorticity sites. For this to happen the
zero-vorticity line must leave the point $z$ on its right. The
probability of such an event depends only on the angle $\theta$
between the point and the origin and it assumes a particularly
simple form in terms of hypergeometric functions$^{27}$.  In the
inset of Figure~\ref{fig:3} we show that the analytic solution
fits very well the numerical data with $\kappa=5.9$. The second
example
 is the probability that a vorticity cluster spans
the rectangle joining two opposite sides. What is the average
number of such spanning clusters?  What is the probability that a
``four-legged cluster'' joins all four sides? By scale invariance
these quantities depend only on the aspect ratio $r$ of the
rectangle, and their precise dependence can be found by exploiting
conformal invariance.  In the context of critical percolation
formulae for such probabilities have been derived %by means of CFT
by Cardy$^{28}$  and Watts$^{29}$ and later proven by
Smirnov$^{30}$  and Dub\'edat$^{31}$. In the main frame of
Figure~\ref{fig:3} we show that numerical data for vorticity
clusters follow very closely the expectations for SLE$_6$. We have
also checked (green symbols in Fig~\ref{fig:4}) that the dimension
of the set of narrow necks that enclose large fjords or large
peninsulae has dimension$^{32}$ $3/4$ (the set is defined by the
pairs of points on the curve that are closer than $L_f$ yet
separated by an arclength larger than $1000\,L_f$). All that gives
further support to the result that zero-vorticity lines are
conformal invariant and belong to the same class of universality
as boundaries of percolation clusters.

Whether the statistics of the zero-vorticity isolines indeed falls
into the simplest universality class of critical phenomena (and
the fractal dimensions are exactly $7/4$ and $4/3$) deserves to be
a subject of more study. Do our findings signify that universal
nature of percolation extends to turbulence as well as to
diffusion-limited aggregation$^{33}$ and quantum chaos$^{34}$? At
the present level it has the status of a tantalizing conjecture
with strong -- though not conclusive -- support from the data. In
view of the non-local constraint imposed by the flow
incompressibility, it is quite surprising that the statistics of
zero-vorticity isolines (within experimental accuracy) enjoys the
locality property inherited by its SLE$_6$ nature. Remind that
continuous percolation can be constructed as a ``flooded
landscape'' determined by some short-correlated random height
function. However, vorticity field in the inverse cascade is not
short-correlated, it has power-law correlation
$\langle\omega(0)\omega({\bf r})\rangle\propto r^{-4/3}$. When the
pair correlation function falls slower than $r^{-3/2}$ then the
system is not expected generally to belong to the universality
class of uncorrelated percolation and even be conformal
invariant$^{35}$. Indeed, we have seen that the field having the
same pair correlation function as the vorticity yet randomized
phases of the Fourier harmonics does not have conformal invariant
isolines (pink symbols in the lower inset in Fig~\ref{fig:2}). We
thus conclude that there is indeed something special about the
vorticity (which has nontrivial phase correlations and higher
moments) produced by 2d turbulence. It is also intriguing to
notice that conformal field theory of critical percolation
possesses a field of scaling dimension 2/3, identical to the one
for the vorticity in Kolmogorov-Kraichnan phenomenology. We may
also wonder how conformal invariance is broken in statistical
properties of non-zero vorticity isolines. Let us stress that we
have found conformal invariance for zero-vorticity isolines, not
yet for correlation functions as envisaged by Polyakov$^3$; in the
related problem of passive scalar in turbulent flow correlation
functions are not conformal invariant$^{6,36}$.

To conclude, we have developed a numerical tool for testing
conformal invariance in physical systems, established this
symmetry (within experimental accuracy) for 2d inverse cascade and
used it as a powerful new tool in turbulence study which allowed
us to make new quantitative predictions confirmed by the
experiment. That shows how conformal invariance spans the whole
physics, from exalted subjects like string theory and quantum
gravity, via statistical mechanics and condensed matter, down to
earthly atmospheric turbulence. 
{\bf Acknowledgements} This work was supported by the grants from
the European network and Israel Science foundation. G.F. thanks A.
Zamolodchikov, A. Polyakov, E. Bogomolny and K. Gawedzki for
useful discussions.

\noindent {\bf Author Information} Reprints and permissions
information is available at npg.nature.com/reprintsandpermissions.
The authors declare that they have no competing financial
interests. Correspondence and requests for materials should be
addressed to G.F. (gregory.falkovich@weizmann.ac.il).
%%%%%%%%%%%%%%%%%%%%%%%%%%%%%%%%%%%%%%%%%%%%%%

\newpage

\newpage

\newpage

\newpage

\newpage

%%%%%%%%%%%%%%%%%%%%%%%%%%%%%%%%%%%%%%%%%%%%%%%%%%%%%%%%%%%%
\newpage

\begin{table*}[!h]
\begin{tabular}{ccccccccccccc}

$N$ & $\mathrm{dx}$ & $\nu$ & $\alpha$ & $u_{rms}$ &

$L_{f}$ & $\ell_{d}$ &

$\varepsilon_{I}$ & $\varepsilon_{\nu}$ & $\varepsilon_{\alpha}$ &

$\eta_{I}$ & $\eta_{\nu}$ & $\eta_{\alpha}$ \\ \hline

%%%%%%%%%

$2048$ & $4.9 \times 10^{-4}$ & $2\times10^{-5}$ & $0.015$ & $0.26$ &

$0.01$ & $2.4\times 10^{-3}$ &

$3.9\times10^{-3}$ & $1.8\times10^{-3}$ & $2.1\times10^{-3}$ &

$39.3$ & $38.0$ & $1.3$ \\

%%%%%%%%%

$4096$ & $2.4\times 10^{-4}$ & $5\times10^{-6}$ & $0.024$ & $0.26$ &

$0.01$ & $1.2\times 10^{-3}$ &

$3.9\times10^{-3}$ & $0.7\times10^{-3}$ & $3.2\times10^{-3}$ &

$39.3$ & $36.1$ & $3.2$ \\

%%%%%%%%%

$8192$ & $1.2 \times 10^{-4}$ & $2\times10^{-6}$ & $0.025$ & $0.27$ &

$0.01$ & $7.8 \times 10^{-4}$ &

$3.9\times10^{-3}$ & $0.3\times10^{-3}$ & $3.6\times10^{-3}$ &

$39.3$ & $35.3$ & $4.0$ \\

%%%%%%%%%

$16384$ & $0.6 \times 10^{-4}$ & $1\times10^{-6}$ & $0.0$ & $0.24$ &

$0.01$ & $5.5 \times 10^{-4}$ &

$3.8\times10^{-3}$ & $0.2\times10^{-3}$ & $3.6\times10^{-3}$ &

$39.5$ & $37.6$ & $1.9$ \\ \hline

\end{tabular}

\caption{Parameters of the simulations. $N$ spatial resolution,
$\mathrm{dx}$ grid spacing, $\nu$ viscosity, $\alpha$ friction,
$L_{f}$ forcing scale, $\ell_{d}=\nu^{1/2}/\eta_{\nu}^{1/6}$
enstrophy dissipative scale, $\varepsilon_{I}$ energy injection
rate, $\varepsilon_{\nu}$ viscous energy dissipation rate,
$\varepsilon_{\alpha}$ energy dissipation by large-scale friction
(energy growth rate for $N=16384$), $\eta_{I}$ enstrophy injection
rate, $\eta_{\nu}$ viscous enstrophy dissipation rate,
$\eta_{\alpha}$ enstrophy dissipation by friction (enstrophy
growth rate for  $N=16384$).} \label{table}

\end{table*}


\vskip 0.3truecm
\begin{thebibliography}{90}


\bibitem{Kra67} Kraichnan, R.~H., Inertial ranges in two-dimensional
turbulence.
{\em Phys.\ Fluids} {\bf 10}, 1417--23 (1967).

\bibitem{KM80} Kraichnan, R.H. \& Montgomery, D.,
Two-dimensional turbulence. {\em Rep. Prog. Phys.} {\bf 43}
547--619 (1980).

\bibitem{Pol2} Polyakov, A. M.,
The theory of turbulence in two dimensions. {\it Nucl. Phys.} {\bf
B396}, 367--85 (1993).

\bibitem{Tab}Tabeling, P., Two-dimensional turbulence: a physicist approach.
{\em Phys.\ Rep.} {\bf 362}, 1--62, (2002).

\bibitem{KG} Kellay, H. \& I. Goldburg, W.,
Two-dimensional turbulence: a review of some recent experiments.
{\em Rep. Prog. Phys.} {\bf 65}, 845--894 (2002).

\bibitem{FGV} Falkovich, G., Gawedzki K., \& Vergassola, M.
Particles and fields in fluid turbulence. {\it Rev. Mod. Phys.}
{\bf73}, 913--975 (2001).

\bibitem{EREC} Chen, S. et al
% R. E. Ecke, G. L. Eyink, M. Rivera, M. Wan, and Z.Xiao,
On vortex-merger and vortex -thinning in a 2D inverse energy
cascade. {\it  57th APS Meeting of the Division of Fluid
Dynamics},
 abstract MK.009 (Seattle, Washington 2004).

\bibitem{BCV} Boffetta, G., Celani, A. \& Vergassola, M.,
Inverse energy cascade in two-dimensional turbulence: Deviations
from Gaussian behavior. {\em Phys. Rev. E} {\bf 61} R29--R32 (2000)
\bibitem{Polyakov} Polyakov, A. M.,
Conformal symmetry of critical fluctuations, {\it JETP Lett.} {\bf
12}, 381--3 (1970).

\bibitem{BPZ} Belavin, A.A., Polyakov, A.M., \& Zamolodchikov, A.A.,
Conformal field theory. {\it Nucl. Phys. B} {\bf 241}, 333--380
(1984).

%\bibitem{KN} Kager, W. and B. Nienhuis,
%A guide to Stochastic L\"owner Evolution and Its Applications,
%{\it J. Stat. Phys.} {\bf115}, 1149--1229 (2004).

\bibitem{Schramm} Schramm, O.,
Scaling limits of loop-erased random walks and uniform spanning
trees. {\it Israel J. Math.} {\bf 118}, 221--288 (2000).


\bibitem{Law1}
 Lawler, G., Schramm, O. \& Werner, W.,
Values of Brownian intersection exponents I: Half-plane exponents,
{\it Acta Math.} {\bf187} 237--273 (2001).
\bibitem{Law2}
 Lawler, G., Schramm, O. \& Werner, W.,Values of Brownian intersection exponents II:
Plane exponents, {\it Acta Math.} {\bf187} 275--308 (2001).
\bibitem{Law3}
 Lawler, G., Schramm, O. \& Werner, W.,Values
of Brownian intersection exponents III: Two-sided exponents, {\it
Ann. Inst. H. Poincare} {\bf38} 109--123 (2002).
\bibitem{Law4}
 Lawler, G., Schramm, O. \& Werner, W.,Conformal
restriction properties. The chordal case. {\it Journal Amer. Math.
Soc.}{\bf 16} 915-955 (2003).
%  The Dimension of the Planar Brownian Frontier is 4/3
%\texttt{arXiv: math.PR/0010165}
\bibitem{Lbook} Lawler, G.,
%Stochastic L\"owner Evolution. \texttt{http://www.math.cornell.edu/$\sim$lawler}
Conformally invariant processes in the plane. {\it Mathematical
Surveys and Monographs} {\bf 114}, 1--242 (2005)


\bibitem{GK} Gruzberg I. \& Kadanoff, L.,
The Loewner equation: maps and shapes. {\it J. Stat. Phys.} {\bf
114} 1183--1198 (2004).

\bibitem{Cardy} Cardy, J.,
SLE for theoretical physicists. \texttt{arXiv:cond-mat/0503313}.

\bibitem{BB1}  Bauer, M. \& Bernard, D.,
Loewner chains. \texttt{arXiv:cond-mat/0412372}.


\bibitem{Loew} L\"owner, K.,
Untersuchungen \"uber schlichte konforme Abbildungen des
Einheitskreises. {\it Math. Ann.} {\bf 89}, 103--121 (1923).

\bibitem{BB} Bauer, M. \& Bernard, D.,
Conformal field theories of Stochastic Loewner Evolutions.
%\texttt{arXiv: math-ph/0206028};
{\it Commun. Math. Phys.} {\bf 239} 493--521 (2003)
%CFTs of SLEs

\bibitem{SD} Saleur, H. \& Duplantier, B.,
Exact determination of the percolation hull exponent in two dimensions,
{\it Phys. Rev. Lett.} {\bf58}, 2325--2328 (1987).

\bibitem{Beffara} Beffara, V.,
The dimension of the SLE curves. \texttt{ arXiv:math.PR/0211322}

\bibitem{KH} Kondev, J. \& Henley, C.,
Geometrical exponents of contour loops on Random Gaussian
surfaces, {\it Phys. Rev. Lett.} {\bf74}, 4580--3 (1995).

\bibitem{Dup} Duplantier, B.,
Conformally invariant fractals and potential theory. {\it Phys.
Rev. Lett.} {\bf 84} 1363--1367 (2000).

%\bibitem{BC}A. Celani and G. Boffetta, in preparation

\bibitem{MR} Marshall, D. \& Rohde, S.,
Convergence of the Zipper algorithm for conformal mapping.
\texttt{http://www.math.washington.edu/$\sim$marshall/preprints/zipper.pdf}

\bibitem{Schramm2} Schramm, O., A percolation formula.
{\it Elect. Comm. Probab.} {\bf 6}, 115--120 (2001).
%\texttt{arXiv:math.PR/0107096}

\bibitem{Cardy-cross} Cardy, J.,
Critical percolation in finite geometries. {\it J. Phys. A} {\bf
25}, L201--L206 (1992)

%\bibitem{Cardy-cross-2} J. Cardy,
%Linking numbers for self-avoiding loops and percolation:
%application to the spin quantum Hall transition
%{\it Phys. Rev. Lett.} {\bf 84}, 3507 (2000)

\bibitem{Watts} Watts, G.,
A crossing probability for percolation in two dimensions. {\it J.
Phys. A} {\bf 29}, L363--L368 (1996)

\bibitem{Smirnov} Smirnov, S.,
Critical percolation in the plane: conformal invariance, Cardy's
formula, scaling limits. {\it C. R. Acad. Sci. Paris, S\'er. I
Math.} {\bf 333(3)}, 239--244 (2001)

\bibitem{Dubedat} Dub\'edat, J.,
Excursion decompositions for SLE$_6$ and Watts' crossing formula.
\texttt{arXiv:math.PR/0405074}.

%\bibitem{Beffara2} Beffara, V.,
%Hausdorff dimensions for SLE$_6$
%{\it Ann. Prob.} {\bf32} (2004) 2606,
%\texttt{ arXiv:math.PR/0204208}

\bibitem{Nienhuis} Nienhuis, B.,
Exact critical point and critical exponents of O(n) models in two
dimensions. {\it Phys. Rev. Lett.} {\bf 49} 1062--1065 (1982).
\bibitem{MF} Meakin, P. \& Family, F.,
Diverging length scales in diffusion-limited aggregation. {\it
Phys. Rev. A} {\bf 34} 2558--2560 (1986).

\bibitem{BS} Bogomolny, E. \& Schmit, C.,
Percolation Model for Nodal Domains of Chaotic Wave Functions.
{\it Phys. Rev. Lett.} {\bf 88} 114102 (2002).
\bibitem{Weinrib} Weinrib A.,
Long-correlated percolation, {\em Phys. Rev. B} {\bf 29}, 387--95
(1984).

\bibitem{FF} Falkovich, G. \& Fouxon, A.,
Anomalous scaling of a passive scalar in turbulence and
equilibrium. {\em Phys.\ Rev.\ Lett.} {\bf 94}, 214502  (2005).


\end{thebibliography}
\end{document}